\documentclass[epj]{svjour_mod}
\newcommand{\ysl}{\mbox{$y$\hspace{-0.5em}\raisebox{0.1ex}{$/$}}}

\usepackage{epsfig}

\begin{document}
\catcode`\@=11 
\def\lsim{\mathrel{\mathpalette\@versim<}}
\def\gsim{\mathrel{\mathpalette\@versim>}}
\def\@versim#1#2{\vcenter{\offinterlineskip
        \ialign{$\m@th#1\hfil##\hfil$\crcr#2\crcr\sim\crcr } }}
\catcode`\@=12 
\def\NP{\hbox{\tiny NP}}
\def\PT{\hbox{\tiny PT}}
\def\DGE{\hbox{\tiny DGE}}

\title{Perturbative and non-perturbative aspects of
 heavy--quark fragmentation\thanks{Invited talk (E.G.) at the HEP2003
 Europhysics Conference in Aachen, Germany. These proceedings are based on~\cite{CG}.}}
\author{Einan Gardi\inst{1} \and Matteo Cacciari\inst{2}
}                     
%
%
\institute{Institut f{\"u}r Theoretische Physik, Universit{\"a}t Regensburg,
D-93040 Regensburg, Germany \and
Dipartimento di Fisica, Universit{\`a} di Parma, Italy,
and INFN, Sezione di Milano, Gruppo Collegato di Parma }
\date{}
%
\abstract{We describe a new approach to heavy--quark fragmentation which is based
on a resummed perturbative calculation and parametrization of power corrections,
concentrating on the $z \longrightarrow 1$ limit, where the heavy meson carries a large
fraction of the momentum of the initial quark. It is shown that the leading power corrections
in this region are controlled by the scale $m(1-z)$. Renormalon analysis is then used to
extend the perturbative treatment of soft and collinear radiation to the non-perturbative regime.
Theoretical predictions are confronted with data on B--meson production in $e^+e^-$ annihilation.
} 
\maketitle
\section{Introduction}
\label{intro}

The heavy--quark fragmentation function $D(z,m^2,\mu^2)$ is the probability distribution
to produce a heavy meson of a heavy quark. It depends on $z$, the momentum fraction of the meson,
on the quark mass $m^2$ and on the factorization scale $\mu^2$. The fragmentation function has a
formal definition~\cite{CS} as the Fourier transform
\begin{equation}
D(z;\mu^2)\equiv
\frac{1}{2\pi\,z}  \int_{-\infty}^{\infty} \frac{dy_{-}}{y_{-}}\,
\exp(i{p}y/z)\, F(p y;\mu^2),
\label{D_def}
\end{equation}
of the hadronic matrix element of a non-local operator on the light-cone ($y^2=0$):
\begin{eqnarray}
\label{F_def}
&&F(p y;\mu^2) \equiv  \\ \nonumber
&&\frac{1}{4\, N_c}\,\sum_{X}\,
{\rm Tr}\left\{ \langle 0 \vert \ysl \Psi(y)\vert H(p)+X\rangle
\langle H(p)+X\vert \overline{\Psi}(0)\vert0\rangle_{\mu^2}
\right\}.
\end{eqnarray}
Here the final state is composed of the measured heavy meson ($H$) carrying momentum $p$
plus anything else ($X$).

\begin{figure}
\epsfig{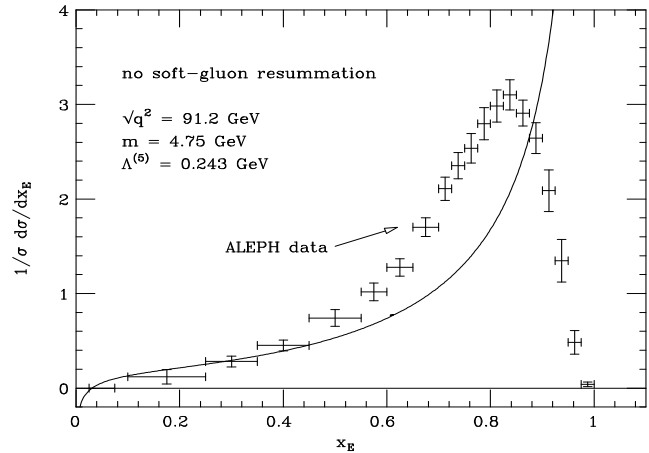}
\caption{ALEPH data for the energy distribution of produced
B mesons at LEP1 compared to the next--to--leading order (NLO)
calculation (full line).\label{nlo}}
\vspace*{-15pt}
\end{figure}

We concentrate here on inclusive observables, the prime example being the single B--meson
inclusive cross section in $e^+e^-$ annihilation, shown in Fig.~\ref{nlo}.
Cross sections of this sort can be written
as a convolution
\begin{equation}
\frac{d\sigma(x,Q^2)}{dx} = \int_x^1\frac{dz}{z}\,{C}(x/z,Q^2;{\mu^2})\,{D}(z;{\mu^2})\,
\label{conv}
\end{equation}
between a process--specific coefficient function $C$, describing the hard
interaction where the heavy quark is produced, and the process--independent
fragmentation function $D$, defined
in~(\ref{D_def}), describing the hadronization stage.

The common practice is to bridge the gap between experimental data
and fixed--order calculations in QCD by means of a fragmentation model,
i.e. a given functional form for ${D}(z;{\mu^2})$ in (\ref{conv})
with one or more free parameters.
For heavy--quark fragmentation the most famous examples are~\cite{kart,peterson}.
Upon excluding the more difficult $z\longrightarrow 1$ region such fits can indeed be performed.
However, the gain is limited: the relation between the parameters
in these models and the matrix elements cannot be made precise.
The models provides no information about the underlying hadronization dynamics. Moreover,
the universality of the extracted parameters is unclear.
At large $z$ fragmentation models simply fail to bridge the gap between the resummed
perturbative calculation and the data. This has been recently demonstrated in a
clear way~\cite{Ben_haim} by directly extracting the ``non perturbative fragmentation component''
from $e^+e^-$ data in moment space and then comparing the resulting distribution in
$z$ space to models.

An important application of the heavy--quark fragmentation function which
demonstrates these problems is in the description of B production in hadron colliders.
The CDF collaboration found~\cite{CDF} an alarming discrepancy
(a factor of 3) between the transverse--momentum distribution of B$^+$ hadroproduction data
and the standard treatment of this cross section, where a NLO calculation is
convoluted with a Peterson model \cite{peterson} for the fragmentation function.
In the latter the free parameter was set to a standard value
based on $e^+e^-$ annihilation data.
Ref.~\cite{CN} applied a resummed perturbative calculation for the coefficient function
and combined it with the relevant fragmentation effect extracted from $e^+e^-$ data
in moment space, concluding that the discrepancy is much smaller.
This shows that the separation between the perturbative
and non-perturbative ingredients of~(\ref{conv}) is very delicate.
A na\"ive application of~(\ref{conv}) simply fails: if the perturbative
ingredient $C$ in (\ref{conv}) is taken at fixed order in $\alpha_s$, the
required ``non-perturbative'' ingredient $D$ appears not to be the same in
different processes.

As heavy--quark production in hadron colliders becomes increasingly important experimentally,
it is evermore urgent to correctly apply perturbative QCD to such cross sections,
to separate in a systematic way between the perturbative and the non-perturbative ingredients,
and finally, to understand hadronization in a quantitative way. In particular, the
parametrization of the fragmentation function $D$ must eventually be understood in
terms of its field theoretic definition (\ref{D_def}).

Our approach to heavy--quark fragmentation is primarily a perturbative one:
we start off with a perturbative calculation of the matrix element in (\ref{F_def}), replacing the
outgoing meson by an on on-shell heavy quark, and treat non-perturbative effects, which make for
the difference between the quark and the meson, as {\em corrections}. Hadronization corrections
are power-suppressed: they are inversely proportional to
the mass of the heavy quark $m$. The perturbative approach is appropriate so long as
$m \gg \Lambda$. Thus it is definitely applicable to bottom, and probably, with some care,
also to charm.

It should be kept in mind that a perturbative calculation is at all possible owing to
two properties: (1) the presence of the quark mass regulating
collinear divergences; and (2) the inclusive nature of the observable, which guarantees
the cancellation of infrared singularities between real and virtual diagrams at
any order in perturbation theory. This cancellation does leave, however, a significant trace
in the expansion: Sudakov logarithms of $1-z$. This is why the
${\cal O}(\alpha_s)$ result shown in Fig.~\ref{nlo} diverges at $z\longrightarrow 1$, whereas
the physical cross section vanishes at this limit. It is only upon summing the $z\longrightarrow 1$
singular terms in the perturbative series to all orders (exponentiation)
that the vanishing of the cross section is recovered.

\section{Asymptotic Scaling}
\label{asym}

Let us first see what can be deduced on the fragmentation function from general considerations.
If the quark mass $m$ is infinitely large, hadronization effects are negligible, and the
fragmentation function is just $\delta(1-z)$. Taking a large but finite ratio $m/\Lambda$,
one would expect the function to be somewhat smeared towards smaller $z$. This smearing
is proportional to $m/\Lambda$, as expressed by the following scaling law
(see e.g.~\cite{Buras:qm}):
$D(z)=(m/\Lambda) f((1-z)m/\Lambda)$.

This property can be formulated more precisely upon taking moments,
\begin{equation}
\tilde{D}(N,m^2) \equiv\int_0^1 dz\; z^{N-1} D(z,m^2),
\label{mom}
\end{equation}
and it can be explicitly derived~\cite{CG} from the field--theoretic definition (\ref{D_def}).
One can consider two limits, one where the mass becomes large and the other where
the moment index~$N$ gets large. For large $m$ one can match the matrix element (\ref{F_def})
onto the heavy--quark effective theory, getting~\cite{JR}:
\begin{equation}
\frac{F(p y, m^2)}{p y}\,\,\exp\left(i{p}y\right)
\longrightarrow \,{{\cal F}(p y \,\bar{\Lambda}/m)}+{\cal O}(\bar{\Lambda}/m),
\label{large-m}
\end{equation}
namely, at the leading order in the large--$m$ expansion
the dependence on $m$ and on the light--cone separation $y_{-}$ (i.e. on $py$) is coupled:
the matrix  element becomes a function of a single argument $p y \,\bar{\Lambda}/m$. Here
$\bar{\Lambda}$ is the difference between the heavy--meson mass $M$ and the heavy--quark mass $m$.
For large $N$ it follows from the definition (\ref{F_def}) and from (\ref{mom}) that
\begin{equation}
\tilde{D}(N,m^2) \longrightarrow \left.{\frac{F(p y,m^2)}{p
y} \exp\left(i{p}y\right)}\right\vert_{py=-iN}\!\!\!\!\!\!\!+\,{\cal
O}\left(\frac1N\right),
\label{large-N}
\end{equation}
namely that to leading order in $1/N$ the $N$-th moment of the fragmentation function
can be obtained by analytically continuing the matrix element as a function of the light--cone
separation to the complex plane and evaluating it at $py=-iN$. From (\ref{large-m}) and
(\ref{large-N}) together it follows that upon taking the simultaneous limit
$m \longrightarrow \infty$ and  $N \longrightarrow \infty$ with a fixed ratio $m/N$,
\begin{equation}
\tilde{D}(N,m^2)\simeq
\left. \,{\cal F}(p y \,\bar{\Lambda}/m) \,
\right\vert_{py=-iN}+{\cal O}\left(\frac1N\right),
\label{large-m-and-N}
\end{equation}
so the fragmentation function becomes a function of a single argument $N\bar{\Lambda}/m$.

In Sec.~\ref{dge} we shall see how the dependence on $m$ and $N$ through the combination
$N\bar{\Lambda}/m$ follows from the large--order behaviour of the perturbative
expansion in the large--$\beta_0$ limit.
Having established Eq.~(\ref{large-m-and-N}) non-perturbatively, we know that this is indeed
the leading behaviour at large $N$ and that corrections to this behaviour
are suppressed by a power of $1/N$.

We see that the scale which characterizes the fragmentation process in the large $z$ region is
$m(1-z)$ or, in moment space,  $m/N$. This scale has a clear meaning when considering the
bremsstrahlung off a heavy quark. Let us examine the emission in a frame where the quark energy
$E$ is much larger than its mass. The radiation pattern (to
${\cal O}(\alpha_s)$) is
\begin{equation}
\frac{dD}{dz\,d\sin^2\theta}\simeq\frac{C_F\,
\alpha_s}{\pi}\,\frac{1}{1-z}\,\frac{\sin^2\theta}{(\sin^2\theta+m^2/E^2)^2},
\label{rad_pattern}
\end{equation}
where only the leading term in the limit $z\longrightarrow 1$ was kept and
the angle of emission $\theta$ is related to the gluon transverse momentum by
$\sin^2\theta={k_{\perp}^2}/\left({E^2z^2(1-z)^2}\right)$. As discussed in~\cite{Dokshitzer:fd},
 the radiation vanishes in the exact forward direction,
 but it peaks close to the forward direction at $\theta\,\simeq\, m/E$  (the `dead cone'),
 or in a boost-invariant formulation at $\vert k_{\perp}\vert \simeq m(1-z)$.
So $m(1-z)$ is the typical transverse momentum of radiated gluons.
The scaling law (\ref{large-m-and-N}) can be understood in physical terms as the observation that
the hadronization effects ($\tilde{D}(N,m^2)$ at large $N$ and $m$) are dominated by
interaction with gluons of transverse momentum $\bar{\Lambda}=M-m$.

\section{Factorization}
\label{fact}

Factorization is based on the fact that dynamical processes taking place on
well--separated physical scales are quantum-mechanically incoherent. This allows one to
treat different subprocess independently of one another and to resum large
corrections.

Eq.~(\ref{conv}) is often regarded as the separation between perturbative and
non-perturbative contributions to the cross section. However, factorization
can be a much stronger tool upon considering separately the dynamics taking place on different
physical scales. Consider, for example, the case of bottom production in $e^+e^{-}$ annihilation,
shown in Fig.\ref{nlo}. Referring to (\ref{conv}) one can na\"ively interpret the gap
between the data and some perturbative calculation as
 the ``non-perturbative fragmentation function''
and then try to bridge this gap using a model. As stressed above this interpretation
leads to much confusion. Instead, the reasons for having large (perturbative and non-perturbative)
corrections need to be identified and the corrections be resummed.

The first step is to separate the scales involved.  Upon neglecting higher order corrections
which are suppressed by powers of $m^2/q^2$ the moments of the cross section can be written
as~\cite{MN,CC}
\begin{eqnarray}
 \tilde{\sigma}(N,q^2,m^2) =
{\tilde{C}(N,q^2;\mu_F^2)}
{\tilde{E}(N,\mu_F^2,\mu_{0F}^2)}
\tilde{D}(N,m^2;\mu_{0F}^2).
\nonumber
\end{eqnarray}
Choosing $\mu_F^2\sim q^2$ and $\mu_{0F}^2\sim m^2$, the coefficient function
$\tilde{C}$ and the fragmentation function $\tilde{D}$ depend only on scales of order
$q^2$ and $m^2$, respectively. The evolution factor $\tilde{E}$ can be obtained solving the
Dokshitzer--Gribov--Lipatov--Altarelli--Parisi (DGLAP)
equation. This factor then resums corrections
depending on $\alpha_s\ln m^2/q^2$ to all orders. Resummation of this kind was implemented
in computing the full line in Fig.~\ref{nlo}. Clearly, this is insufficient.

Next, one observes that the subprocesses $\tilde{C}$ and $\tilde{D}$ may contain additional
large corrections. One generic source of large corrections (see~\cite{Beneke})
are running coupling (renormalon)
effects, which induce factorial growth of the coefficient at high
orders owing to the increasing sensitivity to
extreme ultraviolet or infrared scales. Infrared renormalons in particular are non-summable
and introduce a power--suppressed ambiguity in the perturbative definition of any quantity.
Since for observable quantities this ambiguity must cancel it can serve as a probe
of non--perturbative contributions.

\begin{figure}
\epsfig{file=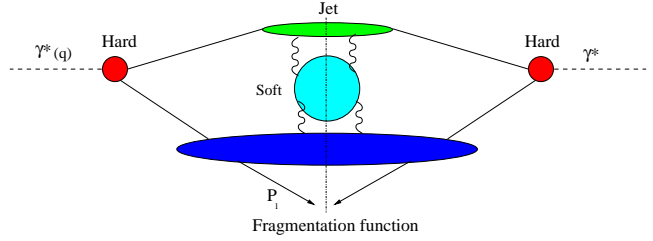,width=8.5cm}
\caption{Factorization of bottom production in $e^+e^-$ annihilation at large $N$.
\label{large_N-fact}}
\vspace*{-10pt}
\end{figure}

Another source of large corrections develops at large $N$: the Sudakov
logs~\cite{Dokshitzer:1995ev,CC}. As stressed above
the fragmentation process is dominated at large $N$ by momenta of order $m/N$. When $m$ and $m/N$
become far apart the concept of factorization applies again, and can be used to resum
logs of $N$ into a Sudakov form factor. This resummation takes the form of exponentiation in
moment space.
A similar situation occurs in the coefficient function $\tilde{C}$, as is demonstrated in
Fig.~\ref{large_N-fact}.
$\tilde{C}$ is
dominated at large $N$ by the invariant mass $q^2/N$ of the unresolved
jet which recoils against the measured heavy meson.
The fact that this jet was also initiated by a heavy quark plays no role at this level~\cite{CG}:
 the relevant scale here is the total invariant mass of the jet.
The same jet function dominates deep inelastic structure functions at
large~$N$~\cite{DIS,Gardi:2002xm}.

It should be emphasized that factorization (contrary to its diagrammatic proofs)
is a non-perturbative concept. One should therefore expect that non-perturbative
corrections on a certain scale would factorise together with the corresponding
perturbative sum. In particular, this must apply to renormalon--related power corrections.
In the case of Sudakov logs factorization leads to exponentiation. Going beyond
the logarithmic level, one finds that power corrections on the corresponding scale
exponentiate as well.
This is the conceptual basis for the ``shape function'' approach to hadronization corrections,
which has been developed in the context of event--shape
distributions~\cite{KS,Korchemsky:2000kp,Gardi:2001ny} (see also \cite{DW}).
This is also the basis of the approach of~\cite{DIS,Gardi:2002xm} to higher twist
in deep inelastic structure functions at large $N$ and of
our approach~\cite{CG} to heavy--quark fragmentation.

\section{Dressed Gluon Exponentiation}
\label{dge}

In order to deal with heavy--quark fragmentation at large~$N$ both Sudakov logs and
renormalons need to be taken into account.
At large $N$, the perturbative coefficients are dominated by Sudakov logs.
However, the resummation of the leading logarithms alone does not provide any information on
power corrections. It is the subleading logs
generated by the running of the coupling which produce the renormalon
ambiguity~\cite{Gardi:2001ny,DGE,CG}. Their resummation is therefore essential to
probe the non-perturbative regime.

From these considerations it follows that the Sudakov exponent needs to be computed to
all orders rather than to some fixed logarithmic accuracy.
Clearly, the full calculation cannot be done. However, relevant all--order
information can be obtained from the large--$\beta_0$ limit corresponding to a
single dressed gluon. Calculating the Sudakov exponent in this way is referred to as
``Dressed Gluon Exponentiation'' (DGE)~\cite{Gardi:2001ny,DGE,CG}.

\begin{figure}                                            %
\epsfig{file=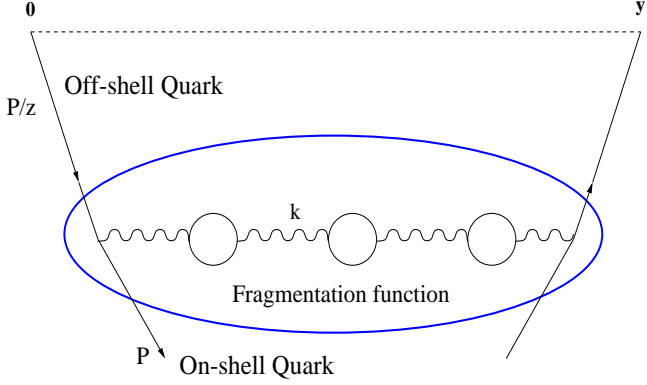,width=8.5cm,height=5cm}
\caption{The diagram contributing to the fragmentation function to leading order in the
flavour expansion (the large--$\beta_0$ limit) in the $A\cdot y=0$ gauge. \label{frag_ren}}
\vspace*{-10pt}
\end{figure}

A process--independent calculation of the fragmentation function~(\ref{D_def}) in the
large--$\beta_0$ limit was performed in~\cite{CG}. In the light-cone axial gauge
$A\cdot y=0$ where the path--ordered exponential is 1, there is just one diagram
-- see Fig.~\ref{frag_ren}.
This diagram was computed using an off-shell gluon splitting
function, which was derived identifying the limit where the massive
quark propagator prior to the emission of the gluon is singular\footnote{In this limit
the gluon virtuality $k^2$, its transverse momentum $k_{\perp}^2$ and the quark mass $m^2$
are taken to be small simultaneously keeping the ratios between them fixed.
This is a generalization of the quasi--collinear limit discussed in~\cite{Catani:2000ef,CC}.}.

The result for the logarithmic derivative of the fragmentation function, written as a scheme
invariant Borel transform, is:
\begin{eqnarray}
\label{dD_dm}
\nonumber
&&\frac{d \tilde{D}(N,m^2)}{d\ln m^2}\!=\! -\frac{ C_F}{\beta_0}\int_0^{\infty}\!\!\!du
{\left(\frac{\Lambda^2}{m^2}\right)}^{u}\!{\rm e}^{\frac53 u}\int_{0}^1 dz \left(z^{N-1}-1\right)
\\
&&  \left(\frac{z}{(1-z)^2}\right)^u  \left[\frac{z}{1-z}\,(1-u)+\frac12(1-z)\,(1+u)\right],
\end{eqnarray}
where $\Lambda$ is in the ${\overline {\rm MS}}$ scheme. A
generalization of this result beyond the large $\beta_0$ limit
which fully captures the next--to--leading logarithms (NLL) was constructed in~\cite{CG}.

Eq.~(\ref{dD_dm}) takes into account the cancellation between real ($z^{N-1}$) and
virtual (1) corrections. In the square brackets we distinguish between $z=1$ singular
and regular terms.
The former lead to logarithmically enhanced contributions in the perturbative expansion, and therefore
need to be exponentiated.

According to (\ref{dD_dm}) the natural scale for the renormalization of the coupling at fixed $z$ is
$(1-z)^2m^2/z$. Thus, integrating over the Borel variable $u$ first is not possible for
$(1-z)m\lsim \Lambda$. As expected, perturbation theory breaks down when the gluon virtuality
or its transverse momentum become comparable to the QCD scale.
This constraint takes a completely different form when considered in moment space: infrared renormalons
show up.

\begin{figure}
\epsfig{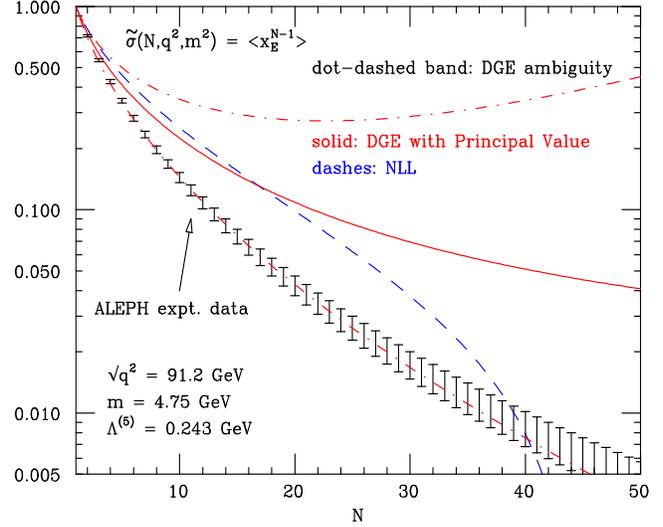}
\caption{The perturbative DGE result compared with ALEPH data for B production
in $e^+e^-$ annihilation, plotted as a function of the moment~$N$.\label{DGE_N}}
\vspace*{-5pt}
\end{figure}

\begin{figure}
\epsfig{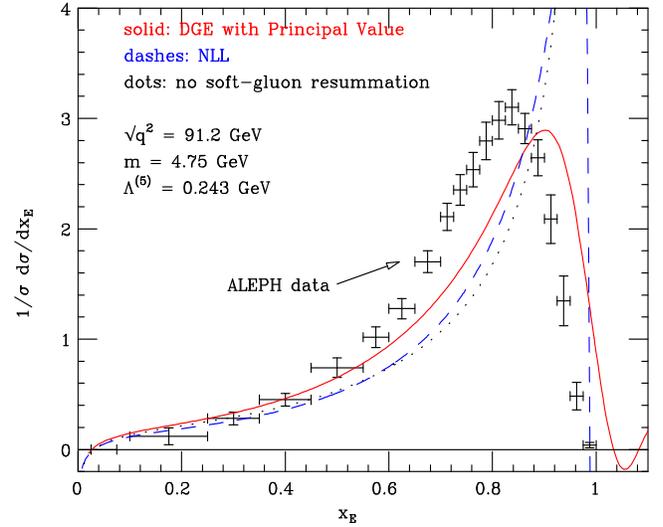}
\caption{Same as Fig.~\ref{DGE_N}, but plotted as a function of the energy fraction of the
measured heavy meson.\label{DGE_x}}
\vspace*{-10pt}
\end{figure}

We proceed to compute the Sudakov exponent in the large--$\beta_0$ limit by isolating
the $z=1$ singular terms, performing the $z$-integration and then integrating over $m^2$.
The result is:
\begin{eqnarray}
\label{D_N_MSbar}
&&\ln \tilde{D}(N,m^2;\mu_{F0}^2)
=\frac{ C_F}{\beta_0}\int_0^{\infty}\!\frac{du}{u}
  \left({\frac{{\Lambda^2}}{m^2}}\right)^{u} \times \\         \nonumber
&&\qquad  \qquad
\bigg[  \left(\frac{m^2}{\mu_{F0}^2}\right)^{u} B_{\cal A}(u) \ln N
-{B_{\tilde{D}}^{\DGE}(u,N)} \bigg],
\end{eqnarray}
where
\begin{equation}
\label{BD}
{ B_{\tilde{D}}^{\DGE}(u,N)}=-\,{\rm e}^{\frac53 u}\,
{ (1-u)}\,\Gamma({-2u})\,\left({N}^{2u}-1\right).
\end{equation}
The $m^2$ integration requires to introduce an ultraviolet subtraction: a
$\mu_{F0}^2$--dependent
counter term which cancels the $u=0$ singularity of the fragmentation function. This term
is the well-known cusp anomalous dimension~\cite{KR,KM}, given by $B_{\cal A}(u)\ln N$, where
$B_{\cal A}(u)=1+\frac53u+\ldots$ (we use the $\overline {\rm MS}$ factorization scheme).
Note that contrary to $B_{\tilde{D}}^{\DGE}(u,N)$
this subtraction term has just a single $\ln N$ to any order in $u$ and it is also free of
infrared renormalon singularities.

According to Eq.~(\ref{BD}), renormalons in the Sudakov exponent (\ref{D_N_MSbar}) appear
at all integer and half integer $u$ values with the exception of $u=1$. It is clear from
Eq.~(\ref{dD_dm}) that these renormalons are exclusively related to the $z\longrightarrow 1$ limit.
To define the perturbative sum corresponding to $\ln \tilde{D}(N,m^2)$ one needs to
integrate over $u$ with some prescription that avoids the poles.
The natural choice is the principal--value (PV) prescription (it was implemented numerically
in \cite{CG}). The ambiguity in choosing a prescription is compensated by
power corrections corresponding to the residues. Introducing a free parameter for each singularity one ends
 up with an additive correction to the perturbative Sudakov exponent having the form:
\begin{equation}
\ln \tilde{D}_{\NP}({ N\Lambda/m})=
-\epsilon_1{\frac{N\Lambda}{m}}-\epsilon_3
\left(\frac{N\Lambda}{m}\right)^3-\epsilon_4
\left(\frac{N\Lambda}{m}\right)^4+\cdots.
\label{NP}
\end{equation}
Finally, exponentiating the result to compute $\tilde{D}(N,m^2)$ the perturbative and
non-perturbative contributions appear as two factors:
\begin{equation}
\tilde{D}(N,m^2;\mu_{F0}^2)={\tilde{D}_{\PT}(N,m^2;\mu_{F0}^2)}\,
\tilde{D}_{\NP}(N\Lambda/m).
\end{equation}
The leading power correction of the form $N\Lambda/{m}$ predicted in \cite{Webber_Nason} is readily
obtained from (\ref{NP}) upon expanding the exponent.

It should be stressed that in
both the perturbative (\ref{D_N_MSbar}) and the non-perturbative (\ref{NP}) contributions
to the fragmentation function we considered here only the leading terms at large $N$.
At the perturbative
level the result is improved \cite{CG} by matching it with the full NLO coefficient.
At the non-perturbative level, there may be additional ${\cal O}(\Lambda/m)$ terms which
we do not parametrize, and consequently the description of the first few moments is of
limited accuracy. In practice, to deal with low moments, it is useful to modify the
parametrization~(\ref{NP}) replacing $N\longrightarrow N-1$ such that the $N=1$ moment
is exactly $1$, as it must be by definition.

The perturbative PV--regulated DGE result of Eq.~(\ref{D_N_MSbar}),
matched to the NLL and the NLO expressions and combined with the proper coefficient function,
is compared as a function of $N$ with the ALEPH data in Fig.~\ref{DGE_N} (full line).
In contrast with the NLL result of Ref.~\cite{CC} (dashed line), the
DGE one does not have a Landau singularity~\cite{CG} and thus it extrapolates
smoothly towards the values of $N\gsim m/\Lambda$ which are beyond perturbative reach.

Also shown in  Fig.~\ref{DGE_N} is the ambiguity
(band shown by two dot-dashed lines) corresponding to the residue of the first renormalon pole located
at $u=1/2$. The lower edge of the band just matches the data, indicating that the power correction of the form
and magnitude(!) expected based on the renormalon analysis is supported by the data.

\begin{figure}
\epsfig{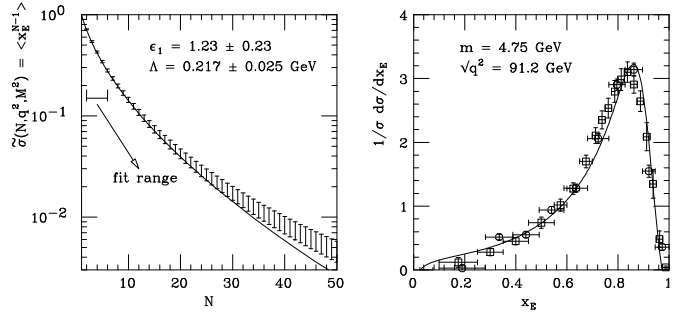}
\caption{A DGE--based fit to ALEPH data. The free parameters are $\alpha_s$ (i.e. $\Lambda$)
and the leading power correction coefficient $\epsilon_1$, which controls the shift of the
perturbative distribution.\label{fit}}
\vspace*{-10pt}
\end{figure}

The different perturbative
results are converted to $x$ space in Fig.~\ref{DGE_x}. Here the significant impact of Sudakov
resummation to NLL as well as that of the additional renormalon resummation achieved by DGE on the shape of
the distribution is evident. Note that the shape of the DGE curve resembles that of the data but it is centered at
larger~$x$. Indeed, the
leading effect of the non-perturbative function (assuming in (\ref{NP}) that only $\epsilon_1\neq 0$)
is a shift of the entire perturbative distribution, very much the same as the leading corrections
in event--shape distributions~\cite{KS,DW,Gardi:2001ny}. Finally, regarding the non-perturbative
parameters $\epsilon_n$ as free parameters in a fit, the data can be well described. The result of a
fit in moment space where the only non-perturbative correction is $\epsilon_1$ is
shown in Fig.~\ref{fit}.

Upon using more non-perturbative parameters the details of the prediction (\ref{NP}) can be confronted with data.
The analysis in \cite{CG} shows that subleading non-perturbative corrections at the exponent
are rather small, and the absence of a correction of the form $N^2\Lambda^2/m^2$
can be consistent with the data.

\section{Conclusions}
\label{concl}

We described here a new approach to the QCD description of heavy--quark fragmentation
concentrating on the $z\longrightarrow 1$ limit.
It was first rigorously demonstrated that the non-perturbative dynamics is dominated by
the scale $m(1-z)$. This scale corresponds in perturbation theory to the transverse momentum of
gluons radiated from the heavy quark.
Based on a renormalon analysis we extended the perturbative technique
for resumming soft gluon radiation to the non-perturbative regime, identified
power--like effects and separated them from the perturbative fragmentation function
by means of a PV prescription. The non-perturbative contribution was then parametrized
based on the renormalon ambiguity.
We found that the simplest possible parametrization of power corrections which follows from
renormalons, namely a shift of the perturbative distribution, is sufficient to
describe the data on B production in $e^+e^-$ annihilation.
This way phenomenological models for the non-perturbative fragmentation function are not needed.

The fragmentation function was treated, based on its definition~(\ref{D_def}), in a process
independent way.
The results are thus applicable independently of the production process, given that the
corresponding coefficient function in the ${\overline{\rm MS}}$ scheme is known.
Universality of the leading power corrections at large $z$ can now be tested experimentally.

\end{document}